\documentclass[epj,twocolumn]{webofc}
\usepackage[varg]{txfonts}   
\woctitle{ICNFP17} 
\newcommand{\beq}{\begin{equation}}
\newcommand{\eeq}{\end{equation}}
\newcommand{\bea}{\begin{eqnarray}}
\newcommand{\eea}{\end{eqnarray}}
\newcommand{\eps}{\varepsilon}

\newcommand{\mkrm}[1]{}           

\begin{document}

\title{Self-consistent account for phonon induced corrections to quadrupole moments of odd
nuclei. Pole and non-pole diagrams.}

\author{E.\,E. Saperstein\inst{1}, \inst{2} 
\and   S. Kamerdzhiev\inst{1} \and   D.\,S. Krepish\inst{1}  \and   S.\,V. Tolokonnikov\inst{1},
\inst{3} \and   D. Voitenkov\inst{4}}

\institute{National Research Centre ``Kurchatov Institute'', 123182, Moscow, Russia \and National
Research Nuclear University MEPhI, 115409 Moscow, Russia \and  Moscow Institute of Physics and
Technology, 141700, Dolgoprudny, Moscow Region, Russia \and Institute for Physics and Power
Engineering, 249033 Obninsk, Russia}

\abstract{Recent results  of the description of quadrupole moments of odd semi-magic nuclei are
briefly reviewed. They are based on the self-consistent theory of finite Fermi systems with account
for the phonon-particle coupling (PC) effects. The self-consistent model for describing the PC effects
was developed previously for magnetic moments. Account for the non-pole diagrams is an important
ingredient of this model.  In addition to previously reported results for the odd In and Sb isotopes,
which are the proton-odd neighbors of even tin nuclei, we present new results for odd Bi isotopes, the
odd neighbors of even lead isotopes. In general, account for the PC corrections makes the agreement
with the experimental data significantly better.}

\maketitle

\section{Introduction}
Spartak Belyaev was one the creators of the microscopic nuclear physics.  His famous article of 1959
\cite{Belyaev-59} contained several cornerstones of modern nuclear theory.  A crucial role of the
first 2$^+$ excitations in even-even spherical nuclei, the quadrupole ``phonons'', is one of them. The
quadrupole phonons are the surface vibrations, belonging to the Goldstone branch related to the
spontaneous breaking of the translation symmetry in nuclei. They play the main role in the problem of
the phonon-particle coupling (PC) corrections to characteristics of the ground states of odd nuclei we
consider.

Consideration of such effects within the self-consistent theory of finite Fermi systems (TFFS)
\cite{scTFFS} or any other approach based on the use of phenomenological parameters is a rather
delicate problem as these parameters could include implicitly some of these effects. In principle, two
strategies can be used. The first one was chosen in  the self-consistent TFFS \cite{scTFFS}. This
approach is based on the general principles of the TFFS \cite{AB} supplemented with the TFFS
self-consistency relation \cite{Fay-Khod}. In this method, the parameters from the beginning were
chosen in order to describe the nuclear characteristics with account for the PC contributions. Another
way was chosen by Fayans with coauthors who developed the energy density functional (EDF) method
\cite{Fay1,Fay4,Fay5,Fay} supposing that the EDF parameters include all the PC effects on average.
Indeed, the Fayns EDFs FanDF$^0$ \cite{Fay5} or DF3 \cite{Fay4,Fay} described the binding energies and
radii of spherical nuclei with rather high accuracy at the mean field level. Later, again at the mean
field level, the EDF DF3 and its version DF3-a \cite{DF3-a} with changed spin-orbit and effective
tensor parameters were successfully applied to describe magnetic \cite{mu1,mu2} and quadrupole
\cite{BE2,Q-EPJ,Q-EPJ-Web} moments of odd spherical nuclei. The characteristics of the first 2$^+$
levels in even Sn and Pb isotopes \cite{BE2,BE2-Web} were also described perfectly well, much better
than in the analogous calculations \cite{BE2-HFB} with the use of the Skyrme EDFs SLy4 and SkM*. The
list of successful applications of the Fayans EDFs to spherical nuclei can be supplemented with
description of the single-particle (SP) spectra of magic nuclei \cite{Levels}. Recently, the Fayans
method with the EDF FaNDF$^0$ was developed for deformed nuclei \cite{Fay-def}. The first applications
of this EDF in this field \cite{Fay-def1,drip-2n,Pb-def,U-def,alpha-def} turned out to be rather
successful.

However, there are PC corrections to characteristics of odd nuclei which behave in a non-regular way
depending on the nucleus under consideration and the SP state $|\lambda\rangle$ of the odd nucleon. At
first, it concerns the PC corrections induced by the $2^+_1$ phonons which are, as a rule, the lowest
excitations of the even-even spherical nuclei. It occurs because some PC corrections to different
nuclear characteristics depend on the $L$-phonon excitation energy $\omega_L$ at small $\omega_L$
values as $1/\omega_L$ \cite{scTFFS}. The excitation energy of the $2^+_1$ phonon behaves usually in a
non-regular way, especially in vicinity of magic nuclei. For example, in the lead isotopic chain we
have $\omega_2{\simeq}4\;$MeV in $^{208}$Pb and $\omega_2{<}1\;$MeV in all lighter even Pb isotopes.
Evidently, it is practically impossible to describe so non-regular behavior of the $L$-phonon
characteristics and the corresponding PC corrections with a universal set of the EDF parameters. Thus,
if we want to reach higher accuracy in reproducing nuclear data, we should try to separate some
fluctuating part of the PC corrections in order to add them to the mean field predictions. Such a
programme was carried out for SP levels in magic \cite{Levels} and semi-magic \cite{semi-Levs} nuclei
and for the double odd-even mass differences of magic \cite{DMD1,DMD2} and semi-magic \cite{DMD3}
nuclei. In all the cases, inclusion of the PC corrections made agreement with the data better.

In Refs. \cite{EPL_mu,Lett_mu,YAF_mu}, a model was developed  to find the fluctuating part of the PC
corrections to magnetic moments of odd semi-magic nuclei found previously within the mean field theory
\cite{mu1,mu2}. A semi-magic nucleus contains two subsystems with absolutely different properties. One
of them is superfluid, whereas the second subsystem is normal. The model under discussion was
developed for nuclei with the odd nucleon belonging to the normal sub-systems. It simplifies the
formulas for the PC corrections drastically. Recently, a similar model was developed for the
quadrupole moments of odd semi-magic nuclei \cite{PC-Q}. The odd-proton neighbors of the even tin
nuclei, i.e. the odd isotopes of In and Sb, were considered in this work.

In Sect. 2, we describe briefly the model under discussion, for the case of quadrupole moments. In
Sect. 3 we present briefly the results of \cite{PC-Q} for the PC corrections, induced by the $2^+_1$
phonons,  to quadrupole moments of odd In and Sb isotopes. The new results of similar consideration
for the odd Bi isotopes are given in Sect. 4. At last, Sect. 5 contains  discussions and conclusions.

\section{The model for PC corrections to multipole moments}

In this Section, we present  briefly the formalism used, for the case of quadrupole moments. For
magnetic moments, all formulas are similar and can be found in \cite{EPL_mu,YAF_mu}.  Within the TFFS
\cite{AB}, quadrupole moments of odd nuclei are determined in terms of the diadonal matrix elements
\beq Q_{\lambda}= \langle\lambda| V |\lambda\rangle_{m=j},\label{Q_lam}\eeq $|\lambda\rangle$ being
the state of the odd nucleon, of the normal component  $V$ of the effective field. In magic nuclei the
latter obeys the  RPA-like equation,  \beq  V = V_0+{\cal F} A V, \label{Vef_s} \eeq where $V_0$ is
the external field, ${\cal F}$ is the Landau-Migdal interaction amplitude, and $A$ is the
particle-hole propagator. For the quadrupole moment problem we discuss, one has $V_0({\bf
r}){=}\sqrt{16\pi/5}r^2 Y_{20}({\bf n})(1+\tau_3)/2$. In fact, we deal with superfluid nuclei, and the
RPA eq. (\ref{Vef_s}) should be replaced with the analogous QRPA equation which can be written
symbolically in the same form, but all the terms are now matrices. The explicit form of this equation
can be found in \cite{AB} or, for the case of the Fayans EDF, which we use as a generator of the
self-consistent basis, in \cite{BE2}.

Let us now consider the PC corrections to the matrix element (\ref{Q_lam}) in the field of the
$L$-phonon, the phonon creation amplitude being $g_L(\bf r)$:  $|\lambda\rangle \to
\tilde{|\lambda\rangle}$, $V_0\to \tilde{V_0}$,   $V\to \tilde{V}$, ${\cal F}\to \tilde{{\cal F}}$,
and $A\to \tilde{A}$, with obvious notation. We follow the scheme of \cite{EPL_mu} and limit ourselves
to the $g^2_L$-approximation. We address to \cite{EPL_mu,YAF_mu} for the detailed formalism. All the
PC corrections we consider are illustrated with the diagrams depicted on Figs. 1 -- 3. Usual graphic
elements are used: the solid line denotes the Green function G, the open circle, the vertex $g_L$, the
dashed triangle is the effective field $V$, whereas the wavy line denotes the phonon $D$-function:
\beq D_L(\omega)= \frac 1 {\omega - \omega_L +i\gamma} - \frac 1 {\omega + \omega_L - i\gamma}
\,.\label{DL} \eeq

\begin{figure}
 \hspace{-0mm}
  \includegraphics[width=120pt]{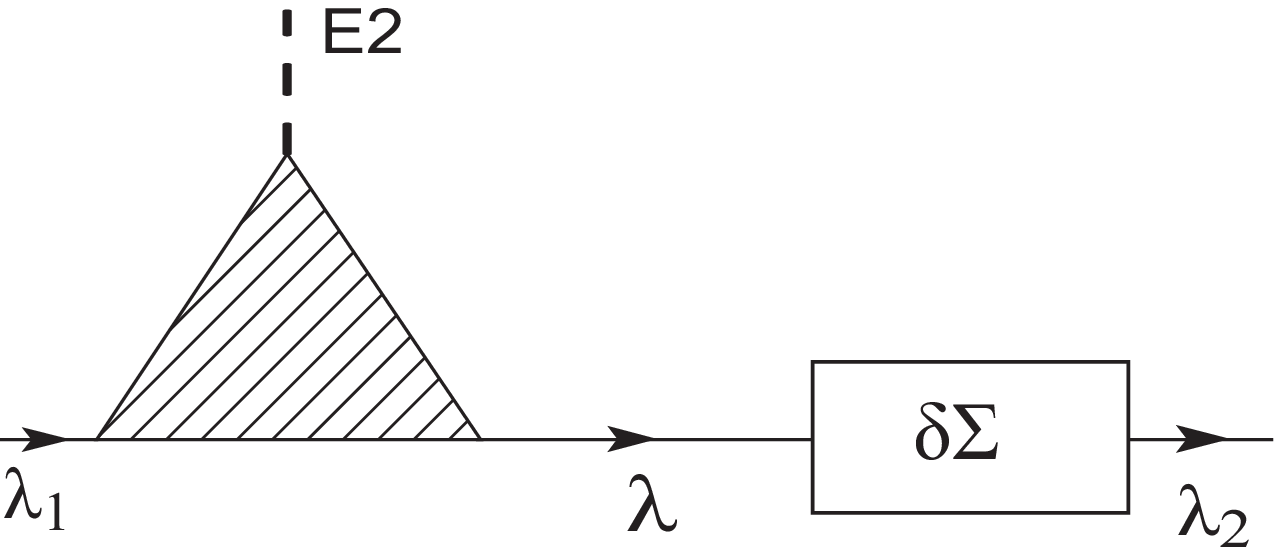}
  \hspace{5mm}
    \includegraphics[width=80pt]{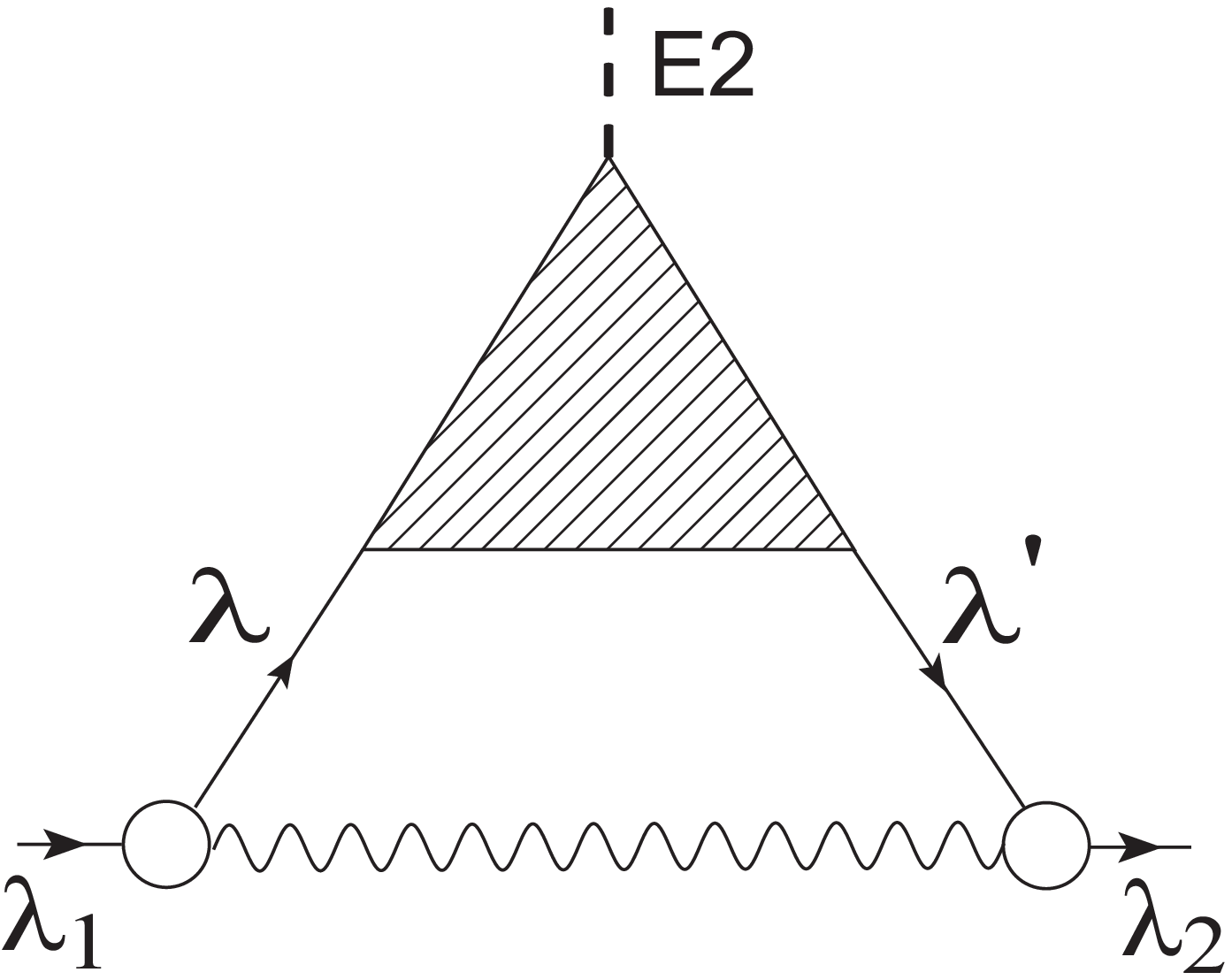}
  \caption{Diagrams for two main PC corrections to the quadrupole moment of an odd nucleus:
  the ``end correction'' (left) and the one due to the induced interaction (right).}
\end{figure}

\begin{figure}
  \centerline{\includegraphics[width=240pt]{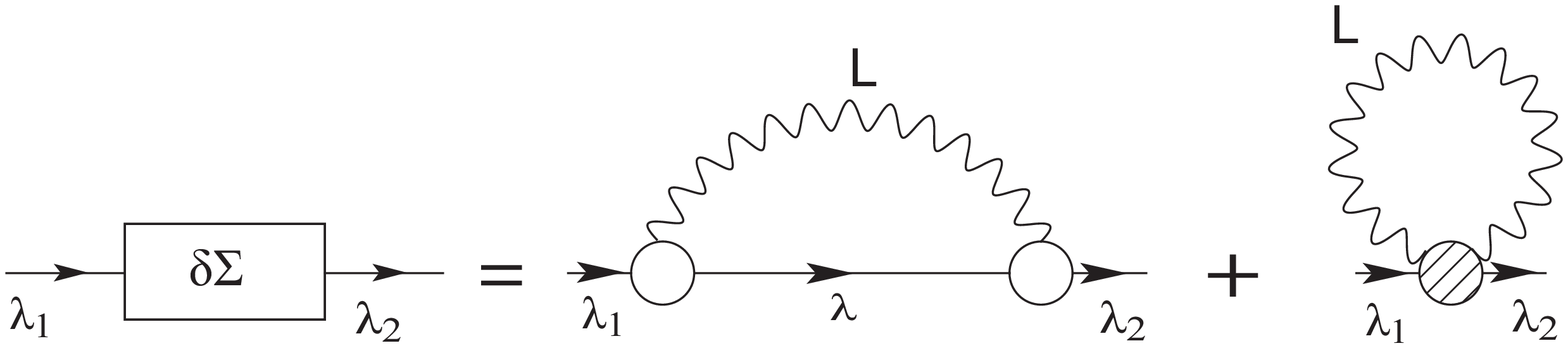}}
  \caption{$g_L^2$ phonon corrections to the mass operator. The dashed blob denotes the sum of non-pole
  (``phonon tadpole'') diagrams.}
\end{figure}

As it was written above, the odd nucleon, the state $|\lambda\rangle$ in (\ref{Q_lam}), belongs to the
normal subsystem of the semi-magic nucleus under consideration. This makes the problem of finding the
PC corrections rather simple. For a superfluid system, an analog of each of the  diagrams in Figs. 1
and 2 looks much more complicated \cite{tad-KS}.

In magic nuclei, the $L$-phonon creation amplitude $g_L$ obeys the homogeneous counterpart of Eq.
(\ref{Vef_s}): \beq g_L(\omega)={\cal F}  A (\omega)  g_L (\omega) \label{g_L}. \eeq In a semi-magic
one, just as in the case of Eq. (\ref{Vef_s}), a matrix QRPA-like analog of this equation appears,
$g\to \hat g_L$, where $\hat g_L$  is the 3-vector containing the normal component $g_L^{(0)}$ and two
anomalous ones, $g_L^{(1,2)}$. As it was demonstrated in \cite{BE2,Q-EPJ}, the normal component is
important only in the problem under consideration. So, the anomalous ones will be neglected and the
upper index will be omitted for brevity, $g_L^{(0)}\to g_L$, in all the formulas below. All the
low-lying phonons we will consider have natural parity, hence
 the vertex $g_L$ possesses  even $T$-parity. It is a
sum of two components with spins $S=0$ and $S=1$, respectively: \beq g_L=  g_{L0}(r) T_{LL0}({\bf
n,\alpha}) +  g_{L1}(r) T_{LL1}({\bf n,\alpha}), \label{gLS01} \eeq where $T_{JLS}$ stand for the
usual spin-angular tensor operators \cite{BM1}. The operators $T_{LL0}$ and $T_{LL1}$ have  opposite
$T$-parities, hence the spin component should be an odd function of the excitation energy,
$g_{L1}\propto \omega_L$. In all the cases we consider, the $2^+_1$ phonon excitation energy is small,
$\omega_2 \sim 1\;$MeV, and the $S=1$ component can be neglected.

Let us begin from the ``end correction'', the left diagram in Fig. 1. Evidently, the analogous diagram
there is, where the left end is PC-corrected.  For the diagonal matrix element (\ref{Q_lam}) of the
effective field, the sum of these corrections corresponds to the following formula: \bea \delta
V_{\lambda\lambda}^{\rm end} & = & - \sum_{\lambda'}V_{\lambda\lambda'}G_{\lambda'}(\eps_{\lambda})
\delta \Sigma_{\lambda'\lambda}(\eps_{\lambda})
   \nonumber \\
& - & \sum_{\lambda'}\delta \Sigma_{\lambda\lambda'}(\eps_{\lambda}) G_{\lambda'}(\eps_{\lambda})
V_{\lambda'\lambda}\label{end1},\eea where \beq G_{\lambda}(\eps) = \frac {n_{\lambda}} {\eps
-\eps_{\lambda} -i \gamma} + \frac {1-n_{\lambda}} {\eps -\eps_{\lambda} +i \gamma}\label{Glam}.\eeq
Two $g_L^2$ diagrams for the PC correction $\delta \Sigma_L$ are depicted on Fig. 2. The second,
non-pole term of $\delta \Sigma$ is rather essential in the problem of the PC corrections to the SP
levels \cite{Levels,semi-Levs}, but it is not so in the problem under consideration
\cite{EPL_mu,YAF_mu,PC-Q}, therefore we do not consider its explicit form here.

On the contrary, the pole term is of primary importance, therefore we write down the corresponding
explicit expression, although it is well-known: \bea \delta\Sigma_{\lambda_2\lambda_1}^{\rm
pole}(\epsilon)&=&\sum_{\lambda\,M}\langle \lambda_2|g^+_{LM}|\lambda\rangle
\langle \lambda|g_{LM}|\lambda_1\rangle \nonumber\\
&\times&\left(\frac{n_{\lambda}}{\eps+\omega_L-
\eps_{\lambda}}+\frac{1-n_{\lambda}}{\eps-\omega_L
-\eps_{\lambda}}\right), \label{dSig2} \eea where
$n_{\lambda}=(0,1)$ stands for the occupation numbers.

After substitution of (\ref{Glam}) and (\ref{dSig2})  into (\ref{end1})  one obtains: \bea  \delta
V_{\lambda\lambda}^{\rm end}& =& - \sum_{\lambda_1 \lambda_2 M} \frac { V_{\lambda\lambda_1}\langle
\lambda|g^+_{LM}|\lambda_2\rangle \langle \lambda_2|g_{LM}|\lambda_1\rangle   } {\eps_{\lambda} -
\eps_{\lambda_1}}
 \nonumber\\ &\times&
 \left(\frac{n_{\lambda_2}}{\eps_{\lambda}+\omega_L-
\eps_{\lambda_2}}+\frac{1-n_{\lambda_2}}{\eps_{\lambda}-\omega_L
-\eps_{\lambda_2}}\right) \nonumber\\ & - & \sum_{\lambda_1
\lambda_2 M}\left(\frac{n_{\lambda_2}}{\eps_{\lambda}+\omega_L-
\eps_{\lambda_2}}+\frac{1-n_{\lambda_2}}{\eps_{\lambda}-\omega_L
-\eps_{\lambda_2}}\right)
 \nonumber\\ &\times&\frac { \langle \lambda|g^+_{LM}|\lambda_2\rangle
\langle \lambda_2|g_{LM}|\lambda_1\rangle V_{\lambda_1\lambda} } {\eps_{\lambda} -
\eps_{\lambda_1}}.\label{end2}  \eea

\begin{figure}
 \hspace{-0mm}
  \includegraphics[width=120pt]{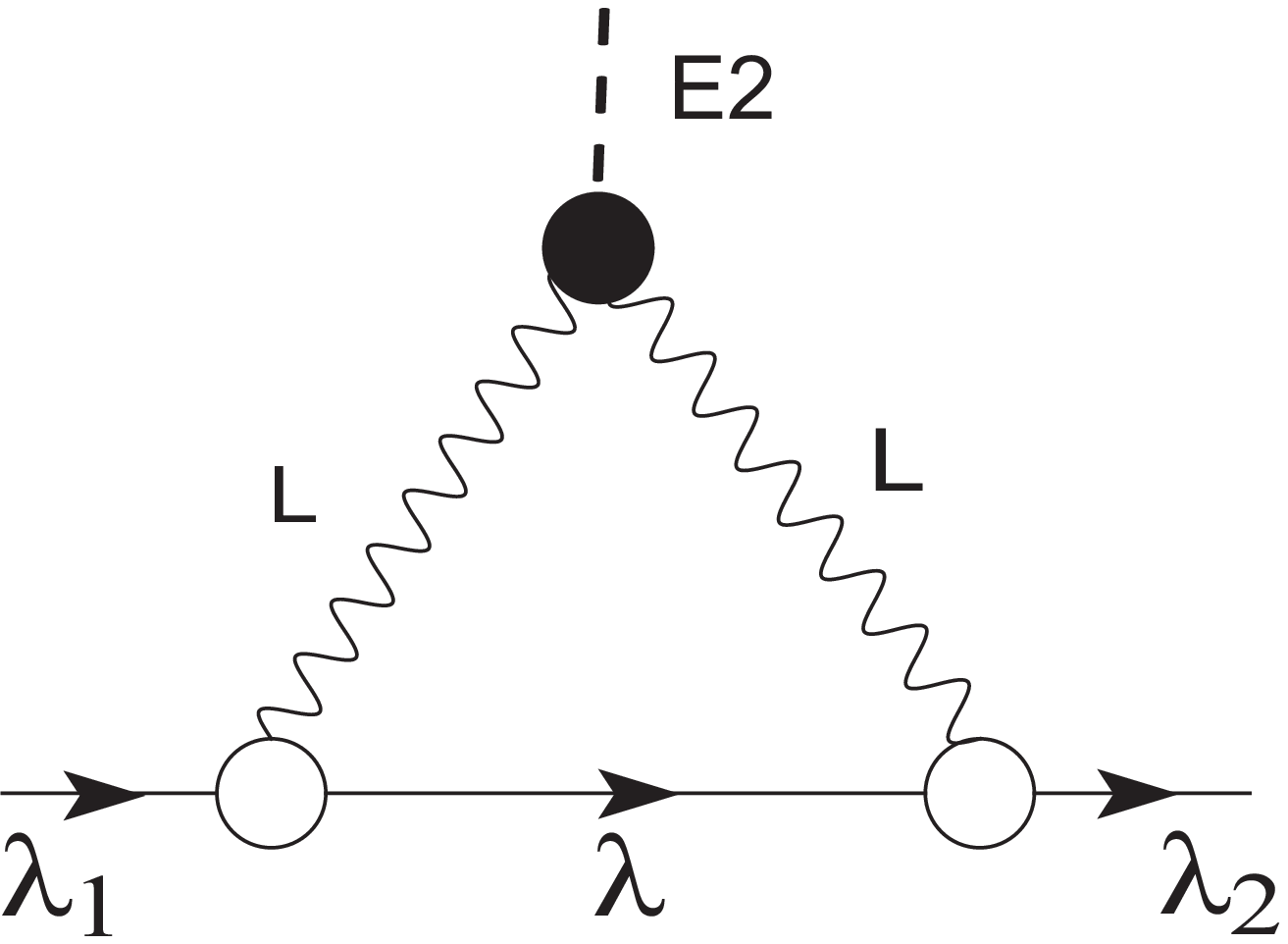}
  \hspace{10mm}
    \includegraphics[width=40pt]{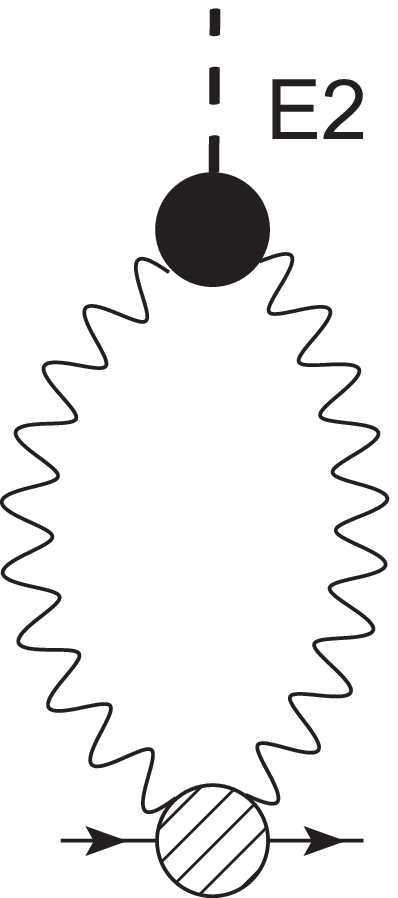}
  \caption{Diagrams for the PC correction due to the quadrupole moment of the $L$-phonon: the triangle
(GDD) diagram (left) and the non-pole one (right).}
\end{figure}

In this equation, the terms with $\lambda_1=\lambda$ in both the sums are singular. This singularity
is removed with the standard renormalization \cite{AB,scTFFS} of the single particle wave functions
$|\lambda\rangle \to \sqrt{Z_{\lambda}} \; |\lambda\rangle$, where \beq Z_{\lambda} = \left( 1- \left.
\frac {\partial \delta \Sigma_{\lambda\lambda}(\eps)}{\partial
\eps}\right|_{\eps=\eps_{\lambda}}\right)^{-1} \label{Zlam} \eeq is the residue of the Green function
at the pole $\eps=\eps_{\lambda}$. This renormalization is the main of the end effects.

However, there is another end effect, which originates from the non-diagonal terms $\lambda_1 \neq
\lambda$ of these sums. It can be calculated directly and is usually rather small
\cite{EPL_mu,YAF_mu,PC-Q}. However, we retain it for completeness, and represent the ``end
correction'' as the sum: \beq  \delta V_{\lambda\lambda}^{\rm end} = \delta V_{\lambda\lambda}^Z +
(\delta V_{\lambda\lambda}^{\rm end})', \label{end3}  \eeq where \beq \delta{V}^Z_{\lambda\lambda} =
\left(Z_{\lambda}-1\right) V_{\lambda\lambda}. \label{endZ} \eeq Note that Eqs. (\ref{end3}),
(\ref{endZ}) correspond  to partial summation of the diagrams of Fig. 1, and, hence, contain  higher
order  terms in $g_L^2$.
 To be consistent up to the order $g_L^2$, the $Z$-factors in
Eqs. (\ref{Zlam}) and (\ref{endZ}) should be expanded in terms of $\partial
\Sigma_{\lambda\lambda}(\eps) /
\partial \eps$, $Z_{\lambda}^{\rm ptb}=1+ \partial \delta \Sigma_{\lambda\lambda}(\eps) /
\partial \eps$, with the result
\beq \left(\delta{V}^Z_{\lambda\lambda}\right)_{\rm ptb} {=}  \left. \frac  {\partial \delta
\Sigma_{\lambda\lambda}(\eps)}{\partial \eps}\right|_{\eps=\eps_{\lambda}}  V_{\lambda\lambda}.
\label{Z_ptb} \eeq The energy derivative of the mass operator (\ref{dSig2}) can be readily found, and
we omit for brevity its explicit expression.

Let us now go to the ``triangle'' diagram ($GGD$) displayed on the right part of Fig. 1. Evidently, it
describes the effect of the induced interaction ${\cal F}_{\rm ind}$ due to the exchange with the
$L$-phonon. Below we write explicitly the corresponding formula for the external field with symmetry
$E2$: \beq \delta V_{\lambda\lambda}^{GGD} =(-1)^{j-m} \!\left(\!\begin{array}{ccc}\! j&\!\!2\!\!&
j\!\\\!-m&\!\! M\!\!& m\!\end{array}\right)\langle \nu \!\parallel \delta V^{GGD}
\parallel \nu \rangle,\label{GGD1}\eeq
where the notation $\lambda=(\nu,m)$  is used, with the reduced matrix element \bea \langle \nu_0
\parallel \delta V^{GGD}
\parallel \nu_0\rangle = \sum_{12} (-1)^{L+j_2-j_1} \left\{\!\begin{array}{ccc} 2&j_0& j_0\\L&
j_2&j_1\end{array}\right\}
  \nonumber\\
 \qquad\qquad \times \langle 2\!\parallel V \parallel\! 1 \rangle
\langle 0\!\parallel \bar g_L \parallel\! 2 \rangle  \langle 1\!\parallel g_L
\parallel\! 0\rangle I_{12}(\omega_L),
 \quad \label{GGD2}\eea
\bea I_{12}(\omega_L)= \frac 1 {\eps_1-\eps_2} \left[\frac {n_1}
{\eps_0-\eps_1+\omega_L} {+}\frac {1-n_1}
{\eps_0-\eps_1-\omega_L} \right.\nonumber\\
\left.  - \frac {n_2} {\eps_0-\eps_2+\omega_L}
  {-} \frac {1-n_2} {\eps_0-\eps_2-\omega_L}
\right],\quad \qquad  \label{GGD3}\eea where $\bar g_L(\omega){=}g_L({-}\omega)$, \beq \bar g_L({\bf
r};\omega){=} g_{L0}(r;\omega) T_{LL0}({\bf n,\alpha})  {-} g_{L1}(r;\omega) T_{LL1}({\bf n,\alpha}).
\label{gLtild} \eeq In Eqs. (\ref{GGD2}), (\ref{GGD3})  the short notation is used: $1=\nu_1$,...

Let us now turn  to the diagrams of Fig. 3 describing the contribution of the phonon quadrupole
moment. In both of them, the black blob means the phonon quadrupole moment $Q_L^{\rm ph}$. The left
one is the usual ($GDD$) triangle, whereas the right diagram is a non-pole counterpart of the left
one, similar to the second diagram in Fig. 2. Contrary to the latter, the  non-pole diagram in Fig. 3
plays the crucial role in the problem under consideration. After separating the angular variables in
the expression for the triangle $(GDD)$, the left diagram in Fig. 3, we obtain \beq \delta
V_{\lambda\lambda}^{GDD} =(-1)^{j-m} \!\left(\!\begin{array}{ccc}\! j&\!\!2\!\!& j\!\\\!-m&\!\! M\!\!&
m\!\end{array}\right)\langle \nu \!\parallel \delta V^{GDD}
\parallel \nu \rangle,\label{GDD1}\eeq
with the reduced matrix element \bea <0 \parallel \delta V^{GDD}
\parallel 0>  = \sum_1 (-1)^{L+j_0+j_1} Q_L^{\rm ph}  \qquad\qquad \nonumber\\
 \times  \sqrt{ \frac {L(L+1)(2L+1)}{4\pi}}
\left\{\!\begin{array}{ccc} j_0&2& j_0\\L& j_1& L\end{array}\right\} \langle 1\!\parallel g_L
\parallel \!0\rangle
\nonumber\\
 \times    \langle 0\! \parallel \bar g_L \parallel\! 1\rangle  \left(I_1^{(1)}
 (\omega_L)+I_1^{(2)}(\omega_L) \right),
  \qquad\qquad \label{GDD2}\eea
\beq I_1^{(1)}(\omega_L) = \frac {1-n_1}{(\eps_0-\eps_1-\omega_L)^2} +  \frac
{n_1}{(\eps_0-\eps_1+\omega_L)^2}, \label{GDD3}\eeq \beq I_1^{(2)}(\omega_L) {=} -\frac 1 {\omega_L}
\left( \frac {n_1}{\eps_0{-}\eps_1{+}\omega_L} + \frac {1-n_1} {\eps_0{-}\eps_1{-}\omega_L} \right).
\label{GDD4}\eeq
 The second integral (\ref{GDD4}) reveals a dangerous behavior at $\omega_L \to 0$.
The non-pole diagram on Fig. 3 possess a similar singularity \cite{YAF_mu}: \beq I^{\rm non-pole}=
\int \frac {d\omega}{2\pi i} D_L(\omega) D_L(\omega-\omega_0)|_{\omega_0\to 0} = \frac {4}{\omega_L}.
\label{tad6}\eeq Its behavior at $\omega_L \to 0$ is just the same, but with the opposite sign, as of
the integral $I_3^{(2)}$, Eq. (\ref{GDD4}). This makes it reasonable to suppose that their sum is
regular at $\omega_L \to 0$.

Let us denote the corresponding terms of (\ref{GDD2}) with (\ref{GDD3}) and (\ref{GDD4}) as
$\delta\overline{} V^{(1),(2)}_{GDD}$. An ansatz was proposed in the model under discussion how to
deal with these two dangerious terms of $\delta V[Q_L^{\rm ph}]$. It was supposed that the  term
$\delta V^{(2)}_{GDD}$ and the tadpole one $\delta V_{\rm tad}$ cancel each other. Such cancelation
does take place for the ``fictitious'' external fields $V_0={\bf j}$ and $V_0=1$, thus providing the
conservation of the total momentum of the system and the total particle number, correspondingly
\cite{AB}. In addition, this is true in the case of the spurious $1^{-}$ phonon. In the result, the
total $g_L^2$ PC correction  to the effective field becomes equal to \beq \delta V= \delta V^Z_{\rm
ptb} + \delta V_{GGD} + \delta V^{(1)}_{GDD}+ \delta V'_{\rm end}. \label{PC-sum} \eeq

The final ansatz for the quadrupole moments with PC corrections is as follows: \beq
\widetilde{V}_{\lambda\lambda} {=} Z_{\lambda} \left( V + \delta V_{GGD} + \delta V^{(1)}_{GDD} +
\delta V'_{\rm end}\right)_{\lambda\lambda}. \label{final} \eeq Just as Eq. (\ref{Zlam}) {\it vs} Eq.
(\ref{Z_ptb}), the ansatz (\ref{final}) differs from the prescription of Eq. (\ref{PC-sum}) in terms
higher in $g_L^2$. It corresponds to making ``fat'' the ends in Fig. 1 and Fig. 3: $|\lambda\rangle
\to \sqrt{Z_{\lambda}} \; |\lambda\rangle$.

\section{PC corrected quadrupole moments of In and Sb isotopes}

In this section, we present the results of application   of the calculation scheme described in the
previous section for the proton-odd neighbors of even Sn isotopes. The self-consistent scheme of
solving Eqs. (\ref{Vef_s}) and (\ref{g_L}) is described in detail in \cite{BE2} and \cite{Q-EPJ}. The
DF3-a version \cite{DF3-a} of the Fayans EDF is used. The excitation energies and quadrupole moments
of $2^+_1$ states in even Sn isotopes, which are ingredients of Eqs. (\ref{end1})--(\ref{final}), are
presented in Table 1. They are taken from \cite{BE2} and \cite{QPRC}. The fresh values of the
experimental quadrupole moments \cite{Stone-2014} are used.

\begin{table}
\begin{center}
\caption{Characteristics of the $2^+_1$ phonons in even Sn isotopes, $\omega_2$ (MeV) and
$Q(2^+_1)$(b)}
\begin{tabular}{|c| c| c| c| c|}

\hline\noalign{\smallskip} $A$  & $\omega_2^{\rm th}$   & $\omega_2^{\rm exp}$
&  $Q^{\rm th}$ &  $Q^{\rm exp}$ \cite{Stone-2014} \\
\noalign{\smallskip}\hline\noalign{\smallskip}
106    & 1.316      & 1.207    & -0.34    &   -    \\
108    & 1.231      & 1.206    & -0.39    &       \\
110    & 1.162      & 1.212    & -0.50    &  0.30(4)      \\
112    & 1.130      & 1.257    & -0.45    & -0.09(10)    \\
114    & 1.156      & 1.300    & -0.28    & (-)0.32(3)        \\
116    & 1.186      & 1.294    & -0.12    & -0.17(4)      \\
118    & 1.217      & 1.230    & -0.01    & -0.14(10)      \\
120    & 1.240      & 1.171    &  0.04    & +0.02(7)      \\
122    & 1.290      & 1.141    &  0.01    & -0.13(10)    \\
124    & 1.350      & 1.132    & -0.07    & +0.03(13)    \\
126    & 1.405      & 1.141    & -0.13   &       \\
128    & 1.485      & 1.169    & -0.14    &       \\

\noalign{\smallskip}\hline

\end{tabular}\end{center}\label{tab1}
\end{table}

Table 2 represents separate PC corrections to quadrupole moments considered in the previous Section.
Here, the $Z$-factor, column 4, is found from its definition (\ref{Zlam}), whereas the perturbation
theory prescription (\ref{Z_ptb}) is used for finding the $\delta Q^Z_{\rm ptb}$ values in column 5.
Thus, the $\delta Q_{\rm ptb}$ quantity, Eq. (\ref{PC-sum}), is just the sum of four partial
corrections in previous columns. At last, the  quantity in the last column is $\delta Q_{\rm
ph}=\tilde{V}-V$ where $\tilde{V}$ is the result of the use of Eq. (\ref{final}) which is the final
prescription of the model we use. We see that two main corrections are those due to the $Z$-factor
(column 5) and due to the induced interaction (the term $\delta Q_{GGD}$, column 6). They always
possess different signs, the sum being significantly less in the absolute value than each of them.
Therefore two other ``small'' corrections are sometimes also important. Note that a technical error
happened in calculations of the quantity $\delta Q_{\rm end}'$ in \cite{PC-Q}, therefore the values in
the column 8 in Table 2 differ from their analogs in \cite{PC-Q}. However, as far as this correction
is very small, all the general results of \cite{PC-Q} remained valid.

It is worth to mention that the $Z$-factor values are often about 0.5 which makes the use of the
perturbation theory in  $g_L^2$ questionable as the value of $(1-Z)$ is a measure of validity of the
$g_L^2$ approximation. Eq. (\ref{final}) we use contains higher in $g_L^2$ terms, but it is just an
ansatz. The analysis shows that the $g_L^2$ approximation in semi-magic nuclei is valid on the
average, but often one ``dangerous'' term appears in Eq. (\ref{dSig2}) with small energy denominator
leading to a big contribution to the value of $(1-Z)$. Simultaneously, the same small denominator
contributes, with opposite sign, to Eq. (\ref{GGD3}) for the induced interaction correction. In the
result, two inaccuracies compensate each other partially. However, a more consistent approach should
be developed for the small denominator situation. Such development  based on the method developed in
\cite{semi-Levs}  is in our nearest plans.

\begin{table*}
\begin{center}
\caption{Various PC corrections, induced by the $2^+_1$ phonon, to the quadrupole moments of
odd-proton In and Sb nuclei. $Q$ is the quadrupole moment without PC corrections \cite{Q-EPJ}. Other
notation is explained in the text. All values, except $Z$ (dimentionless), are in b.}
\begin{tabular}{|l| c| c| c| c| c| c| c| c| c|}
\noalign{\smallskip}\hline\noalign{\smallskip}    nucl.  &$\lambda$ & $Q$ & $Z$ &$\delta Q^Z_{\rm
ptb}$& $\delta Q_{GGD}$ & $\delta Q_{GDD}$ &
$\delta Q_{\rm end}'$ & $\delta Q_{\rm ptb} $  & $\delta Q_{\rm ph} $\\
\noalign{\smallskip}\hline\noalign{\smallskip}
$^{105}$In & $1g_{9/2}$ &  +0.833 &0.675 &-0.400 &0.231 &0.055 & -0.004 &-0.118  &-0.80\\

$^{107}$In & $1g_{9/2}$ &  +0.976 &0.584 &-0.692 &0.404 &0.094 & -0.008 &-0.201  &-0.117\\

$^{109}$In & $1g_{9/2}$ & +1.113 &0.573 &-0.826 &0.487 &0.128 & -0.009 & -0.221 & -0.126\\

$^{111}$In & $1g_{9/2}$ & +1.165 &0.488 &-1.220&0.722 &0.163  & -0.001 & -0.349 & -0.170\\

$^{113}$In & $1g_{9/2}$ & +1.117 &0.576 &-0.820 &0.484 &0.071 & -0.010 & -0.275 & -0.159\\

$^{115}$In & $1g_{9/2}$ & +1.034 &0.609 &-0.662 &0.389 &0.026 &-0.008 & -0.255 & -0.155\\

$^{117}$In & $1g_{9/2}$&   +0.963 &0.632 &-0.560 &0.328 &0.002 &  -0.007 & -0.237 & -0.150\\

$^{119}$In & $1g_{9/2}$&   +0.909 &0.621 &-0.553 &0.323 &-0.008& -0.007 & -0.244 & -0.152\\

$^{121}$In & $1g_{9/2}$&   +0.833 &0.639 &-0.465 &0.271 &-0.002& -0.006 & -0.208 & -0.132\\

$^{123}$In & $1g_{9/2}$&   +0.743 &0.720 &-0.289 &0.168 &0.009 & -0.003 & -0.115 & -0.083\\

$^{125}$In & $1g_{9/2}$&   +0.663 &0.738 &-0.232 &0.134 &0.015 & -0.003 & -0.088 & -0.064\\

$^{127}$In & $1g_{9/2}$&   +0.550 &0.800 &-0.138 & 0.079  &0.012 & -0.001 & -0.049 & -0.039 \\

$^{115}$Sb & $2d_{5/2}$&  -0.882  &0.551 &0.717  &-0.275 &-0.025   & -0.022 & 0.395  & 0.218  \\

$^{117}$Sb & $2d_{5/2}$&  -0.817  &0.582 &0.588  &-0.229 &-0.009    & -0.019 & 0.287  & 0.173 \\

$^{119}$Sb & $2d_{5/2}$&  -0.763  &0.602 &0.504  &-0.198 &-0.001 & -0.019 & 0.285  & 0.169 \\

$^{121}$Sb & $2d_{5/2}$&  -0.721  &0.591 &0.497  &-0.196 &0.003  & -0.017 & 0.240  & 0.145\\

$^{123}$Sb  & $1g_{7/2}$& -0.739  &0.570 &0.552  &-0.328 &0.001  & -0.011 & 0.136  & 0.095\\

\noalign{\smallskip}\hline\noalign{\smallskip}
\end{tabular}
\end{center}
\label{tab:delQ_ph}
\end{table*}

\begin{table*}
\begin{center} \caption{Quadrupole moments $Q\;$(b) of odd In and Sb isotopes.
The theoretical values are $Q_{\rm th}$ and $\tilde{Q}_{\rm th}$ without and with PC corrections,
correspondingly. The differences $\delta Q= Q_{\rm th}-Q_{\rm exp}$ and $\delta
\tilde{Q}=\tilde{Q}_{\rm th}-Q_{\rm exp}$ are given in the last two columns.}

\begin{tabular}{|l |c |c |c |c |c |c|}
\noalign{\smallskip}\hline\noalign{\smallskip}
  nucl.  &$\lambda$  & $Q_{\rm exp}$ &
$Q_{\rm th}$ &$\tilde{Q}_{\rm th}$& $\delta Q$ &$\delta \tilde{Q}$\\
\noalign{\smallskip}\hline\noalign{\smallskip}

$^{105}$In & $1g_{9/2}$& +0.83(5) & +0.83 &  +0.75  &0.00 &-0.08 \\

$^{107}$In & $1g_{9/2}$& +0.81(5) & +0.98 &  +0.86  &0.17 & 0.05   \\

$^{109}$In & $1g_{9/2}$& +0.84(3) & +1.11 &  +0.98  &0.27 & 0.14 \\

$^{111}$In & $1g_{9/2}$& +0.80(2) & +1.17 &  +0.99  &0.36 & 0.19 \\

$^{113}$In & $1g_{9/2}$& +0.80(4) & +1.12 &  +0.96  &0.32 & 0.16\\

$^{115}$In & $1g_{9/2}$& +0.81(5) & +1.03 &  +0.88  &0.22 & 0.07  \\

&                       & 0.58(9)&   &              &0.45 & 0.30 \\

$^{117}$In & $1g_{9/2}$& +0.829(10)& +0.96   & +0.81  &0.136 &-0.02  \\

$^{119}$In & $1g_{9/2}$& +0.854(7) & +0.91   & +0.76  &0.055 &-0.09 \\

$^{121}$In & $1g_{9/2}$& +0.814(11) & +0.83  & +0.70  &0.019 &-0.11 \\

$^{123}$In & $1g_{9/2}$& +0.757(9)  & +0.74  & +0.66  &-0.014 &-0.10 \\

$^{125}$In & $1g_{9/2}$& +0.71(4)   & +0.66  & +0.60  &-0.05  &-0.11 \\

$^{127}$In & $1g_{9/2}$& +0.59(3)   & +0.55  & +0.51 &-0.04   &-0.08  \\

$^{115}$Sb & $2d_{5/2}$& -0.36(6)   & -0.88  & -0.66&-0.52    &-0.30  \\

$^{117}$Sb & $2d_{5/2}$&    -       & -0.817 & -0.63& &-   \\

$^{119}$Sb & $2d_{5/2}$& -0.37(6)   & -0.76  & -0.59  &-0.40 &-0.22 \\

$^{121}$Sb & $2d_{5/2}$& -0.36(4)   &-0.72   & -0.55  &-0.36 &-0.19 \\
&                      & -0.45(3)   &        &       &-0.27 &-0.10 \\

$^{123}$Sb  & $1g_{7/2}$& -0.49(5)  & -0.74 & -0.63  &-0.25 &-0.14 \\

\hline
\end{tabular}
\end{center}
\label{tab:Q_p}
\end{table*}
\begin{figure}
\centerline {\includegraphics [width=70mm]{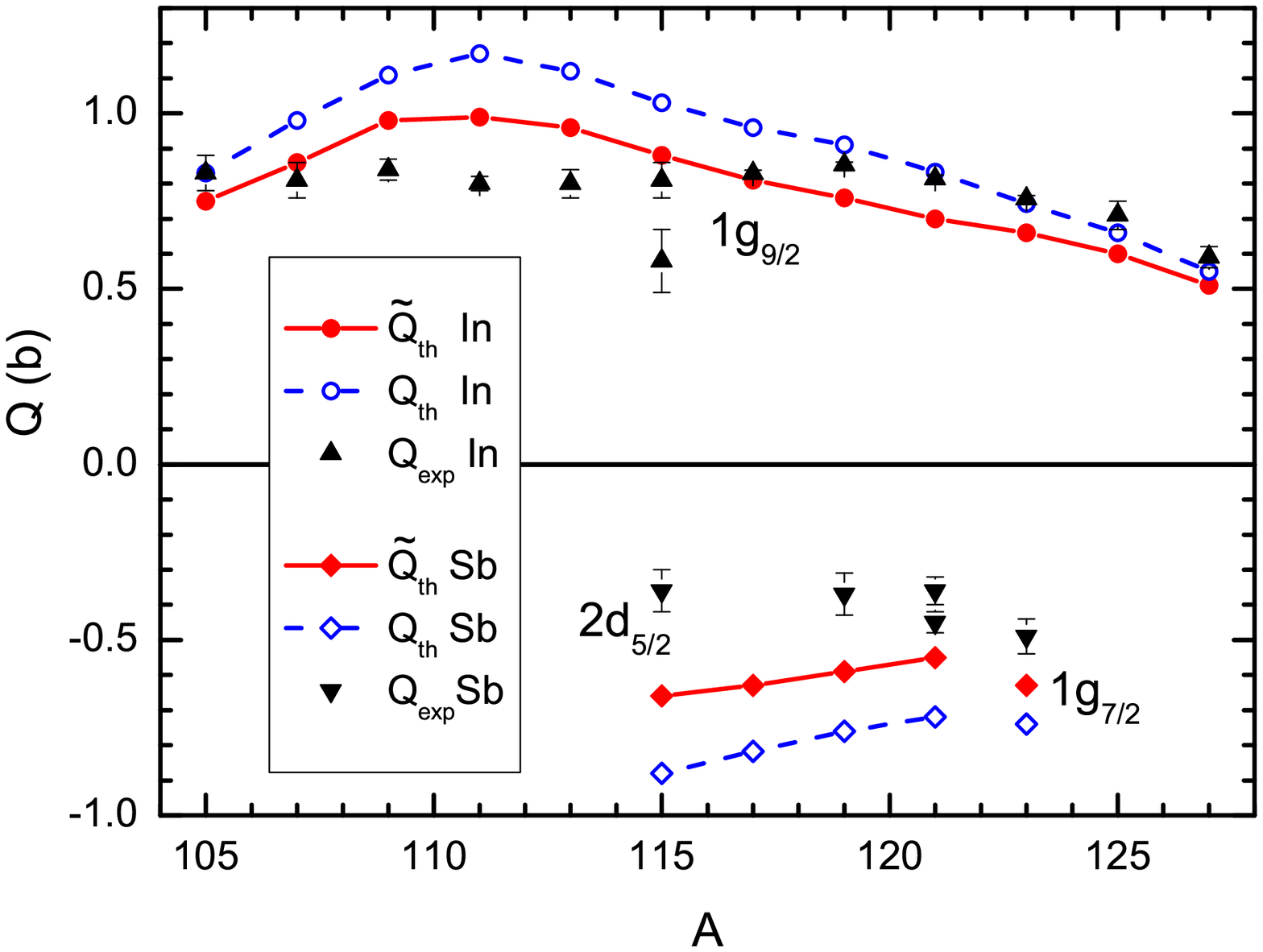}}  \caption{(Color online) Quadrupole moments of odd
In and Sb isotopes with and without PC corrections.}
\end{figure}

The final results are presented in Table 3 and Fig. 4. We see that the PC corrections to quadrupole
moments taken into account make agreement with experiment better in most cases. In any case, this is
true in all the cases where the deviation from experiment of the results  without PC corrections is
significant, more than 0.1 b, which is evidently a typical accuracy of the theory we develop. For
example, this is so for nuclei $^{109,111}$In and for all Sb isotopes. The rms value $<\delta
\tilde{Q}>_{\rm rms}=0.16\;$b follows from the last column of Table 3. The corresponding value without
PC corrections is significantly bigger, $<\delta {Q}>_{\rm rms}=0.27\;$b.

\section{PC corrected quadrupole moments  Bi isotopes}
\begin{table}[h]
\begin{center}
\caption{Characteristics of the $2^+_1$ phonons in even Pb isotopes, $\omega_2$ (MeV) and
$Q(2^+_1)$(b)}
\begin{tabular}{|c| c| c| c| c|}

\hline\noalign{\smallskip} $A$  & $\omega_2^{\rm th}$   & $\omega_2^{\rm exp}$
&  $Q^{\rm th}$ &  $Q^{\rm exp}$ \cite{Stone-2014} \\
\noalign{\smallskip}\hline\noalign{\smallskip}
202    & 0.823      & 0.960    & -0.15    &   -    \\
204    & 0.882      & 0.899    &  0.10    &  +0.23(9) \\
206    & 0.945      & 0.803    &  0.09    &  +0.05(9)        \\
208    & 4.747      & 4.086    &  0.05    &  -0.7(3)      \\

\noalign{\smallskip}\hline
\end{tabular}
\end{center}\label{tab1}
\end{table}

In this Section, the results are presented of new calculations  of the PC corrections, induced by the
$2^+_1$ phonons, to quadrupole moments of four odd $^{203-209}$Bi isotopes, the odd-proton neighbors
of the even $^{202-208}$Pb nuclei. The scheme of presentation of the results is the same as in the
previous Section for the odd-proton neighbors of the even tin isotopes. In Table 4, the
characteristics of the $2^+_1$ states, which are used in our calculations, are given. Tables 5 and 6
present the separate PC corrections and the results for the PC corrected quadrupole moments,
correspondingly. In the Table 4, we see that the excitation energy $\omega(2^+_1)$ value in the magic
$^{208}$Pb nucleus, both experimental and theoretical, is much higher than in the neighboring
semi-magic lead isotopes. Such a behavior of the $2^+_1$ levels is typical for magic nuclei, and the
high value of $\omega(2^+_1)$ is one of the obligatory characteristics of a magic nucleus. So high
value of the excitation energy means that this state possesses with low collectivity. In the result,
the corresponding PC corrections should be very small. The individual role of separate phonons in the
PC corrections to the SP energies of the $^{208}$Pb nucleus was examined in \cite{Levels}. It was
found that the lowest of them, the $3^-_1$ phonon, produces the main contribution to the PC
correction, about 50\%. The other eight phonons (two $5^-$, two $2^+$, two $4^+$, and two $6^+$)
together give the rest of 50\%. Thus, it is obvious from the beginning, that the PC correction to the
quadrupole moment of the $^{209}$Bi nucleus, induced by the $2^+_1$ phonon, will be negligible. Tables
5 and 6 confirm this. Evidently, the PC correction due to the $3^-_1$ phonon in this case should be
much more important.

Other general features of the PC corrections to quadrupole moments found in the previous Section for
In and Sb nuclei also persist here. We mean the main role of the $\delta Q^Z_{\rm ptb}$ and $\delta
Q_{GGD}$ corrections which possess with opposite signs and strongly cancel each other. As to a
comparison with experiment, the data presented in the compilation \cite{Stone-2014} are,
unfortunately, too contradictory for each of four Bi isotopes we consider to make definite
conclusions.

\begin{table*}
\begin{center}
\caption{Different PC corrections, induced by the $2^+_1$ phonon, to the quadrupole moments of
odd-proton Bi nuclei. All notations are the same as in Table 2. }
\begin{tabular}{|l| c| c| c| c| c| c| c| c| c|}
\noalign{\smallskip}\hline\noalign{\smallskip}    nucl.  &$\lambda$ & $Q$ & $Z$ &$\delta Q^Z_{\rm
ptb}$& $\delta Q_{GGD}$ & $\delta Q_{GDD}$ &
$\delta Q_{\rm end}'$ & $\delta Q_{\rm ptb} $  & $\delta Q_{\rm ph} $\\
\noalign{\smallskip}\hline\noalign{\smallskip}
$^{203}$Bi & $1h_{9/2}$ & -1.320  &0.552 & 1.07  & -0.773  & -0.041  &-0.060   & 0.199   &0.110 \\

$^{205}$Bi & $1h_{9/2}$ & -0.954 &0.736 & 0.342 &-0.247   &  0.012  &-0.020  & 0.087   &0.064 \\

$^{207}$Bi & $1h_{9/2}$ & -0.454 &0.867 & 0.070 &-0.051   &  0.005  &-0.005  & 0.019   &0.016  \\

$^{209}$Bi & $1h_{9/2}$ & -0.348 &0.996 & 0.002 & -0.001   & 0.1E-4 & -0.001 & -0.001  &-0.001   \\

\noalign{\smallskip}\hline\noalign{\smallskip}
\end{tabular}
\end{center}
\label{tab:delQ_ph}
\end{table*}

\begin{table*}
\begin{center} \caption{Quadrupole moments $Q\;$(b) of odd Bi isotopes.
All notations are the same as in Table 2.}

\begin{tabular}{|l |c |c |c |c |c |c|}
\noalign{\smallskip}\hline\noalign{\smallskip}
  nucl.  &$\lambda$  & $Q_{\rm exp}$ &
$Q_{\rm th}$ &$\tilde{Q}_{\rm th}$& $\delta Q$ &$\delta \tilde{Q}$\\
\noalign{\smallskip}\hline\noalign{\smallskip}

$^{203}$Bi & $1h_{9/2}$& -0.93(7)  &-1.32  & -1.21  &-0.39  &-0.28  \\
          &           & -0.67(7)  &       &        & -0.65 &-0.54 \\

$^{205}$Bi & $1h_{9/2}$&-0.81(3)  &-0.95  & -0.89   &-0.14  & -0.08 \\
          &           &-0.59(4)  &  &              &-0.36  & -0.30  \\

$^{207}$Bi & $1h_{9/2}$&-0.76(2)  &-0.45  &-0.44    & 0.31  & 0.32 \\
          &           &-0.55(4)       &  &         & 0.10  & 0.11 \\
          &           &-0.60(11)       &  &        & 0.15  & 0.16 \\

$^{209}$Bi & $1h_{9/2}$&-0.516(15)     &-0.348  &-0.349  &0.168  & 0.167  \\
          &           &-0.37(3)       &  &              &0.02  &0.02  \\
         &           &-0.55(1)       &  &              &0.20  &0.20  \\
          &           &-0.77(1)       &  &              &0.42  &0.42  \\
          &           &-0.40(5)       &  &              &0.05  &0.05  \\
          &           &-0.39(3)       &  &              &0.04  &0.04  \\

\hline
\end{tabular}
\end{center}
\label{tab:Q_p}
\end{table*}

\section{Discussion and conclusions}

The results are presented  of the self-consistent  calculations of the PC corrections to quadrupole
moments of odd semi-magic nuclei on the base of the Fayans EDF DF3-a. The self-consistent model is
used, developed previously in \cite{EPL_mu,YAF_mu} for magnetic moments and extended for the case of
quadrupole moments. The main content of  the article consists in the discussion of the results of
\cite{PC-Q} for the PC corrections, induced by the first $2^+$ phonon, in the case of the odd In and
Sb isotopes, the odd-proton neighbors of the even tin nuclei.  In addition, we present the new results
for Bi isotopes, neighboring to the even lead nuclei. The perturbation theory in $g_L^2$ is used,
where $g_L$ is the vertex of creating the $L$-phonon.

The main idea of our approach is to refuse from calculation of all terms proportional to $g_L^2$, as
the main their part is taken into account implicitly in the EDF parameters we use. Instead, only such
$g_L^2$ diagrams are separated and calculated explicitly which behave in a non-regular way, i.e. which
depend significantly on the nucleus under consideration and the state $\lambda$ of the odd nucleon.
Two main PC corrections are the term $\delta Q^Z$ connected with the renormalization of the ends of
all the PC diagrams due to the phonon $Z$-factor and the one, $\delta Q_{GGD}$, due to the
phonon-induced interaction.  They always possess opposite signs and cancel each other significantly.
Therefore two ``small'' corrections are also contribute significantly.

For the calculation of one of them, the term $\delta Q_{GDD}$ due to the quadrupole moment of the
$L$-phonon, account for the non-pole diagram is of primary importance. As we wrote above, the first
$2^+$ phonons were taken into account only in this set of calculations. Their quadrupole moments found
in \cite{QPRC} were used. The sum of all four PC corrections to quadrupole moments of odd In and Sb
isotopes, in most cases, improves agreement with experiment. For the sample of 18 nuclei we consider,
the rms value of the difference between the theoretical predictions and experimental values diminishes
to $<\delta \tilde{Q}>_{\rm rms}=0.16\;$b from the value of $<\delta {Q}>_{\rm rms}=0.27\;$b for the
calculation of \cite{Q-EPJ} without PC corrections.

As to the Bi isotopes, the experimental data \cite{Stone-2014} are, unfortunately, too contradictory
to make definite conclusions about agreement between them and our theory even at the mean-field level.
We hope that the high accuracy of our description of the quadrupole moments of In and Sb isotopes will
be a stimulus for experimentalists to clear up the situation with quadrupole moments of Bi isotopes.

Next step we plan includes the contributions to the PC corrections of the first $3^-$ states. For this
aim, their quadrupole moments for the tin and lead isotopes should be first found within the
self-consistent approach developed in \cite{QPRC}.

\section{Acknowledgments} We acknowledge for support the Russian Science Foundation, Grants Nos.
16-12-10155 and 16-12-10161.

The work was also partly supported  by the RFBR Grant 16-02-00228-a. This work was carried out using
computing resources of the federal center for collective usage at NRC ``Kurchatov Institute'',
http://ckp.nrcki.ru.

EES thanks the Academic Excellence Project of the NRNU MEPhI under contract by the Ministry of
Education and Science of the Russian Federation No. 02. A03.21.0005.

\end{document}